\documentclass[conference]{IEEEtran}
\usepackage{amsfonts}
\usepackage{mathrsfs}
\usepackage{amssymb,amsmath}

\usepackage{algorithm}
\usepackage{algorithmic,mathrsfs}
\usepackage{amsmath,amssymb,epsfig,graphics,subfigure}
\usepackage{theorem}
\usepackage{array,color}
\usepackage[compress]{cite}

\theoremheaderfont{\normalfont\bfseries}

\begin{document}

\title{Robust Linear Transceiver Design for Multi-Hop Non-Regenerative MIMO Relaying Systems}

\author{Chengwen Xing$^\dag$, Zesong Fei$^\dag$, Shaodan Ma$^\ddag$, Jingming Kuang$^\dag$, and Yik-Chung Wu$^\ddag$
 \\ $^\dag$School of Information and Electronics, Beijing Institute of Technology, Beijing, China
\\ Email: \{xingchengwen\}@gmail.com \ \{zesongfei, jmkuang\}@bit.edu.cn \\
$^\ddag$Department of Electrical and Electronic Engineering, The University of Hong Kong, Hong Kong\\
Email: \{sdma,ycwu\}@eee.hku.hk}

\maketitle

\begin{abstract}

In this paper, optimal linear transceiver designs for multi-hop
amplify-and-forward (AF) Multiple-input Multiple-out (MIMO) relaying systems with Gaussian distributed channel estimation errors are investigated. Some commonly used transceiver design criteria are unified into a single matrix-variate optimization problem. With novel applications of majorization theory and properties of matrix-variate function, the optimal structure of robust transceiver is first derived. Based on the optimal structure, the original transceiver design problems are reduced to much simpler problems with only scalar variables whose solutions are readily obtained by iterative water-filling algorithms. The performance advantages of the proposed robust designs are demonstrated by the simulation results.

\end{abstract}

\section{Introduction}
\label{sect:intro}

In order to satisfy the emerging requirements for high speed ubiquitous wireless communications, MIMO cooperative communication has become one of the key parts in the future wireless standards such as LTE, IMT-Advanced, Winner project, etc. Transceiver design for amplify-and-forward (AF) MIMO relaying systems has been reported in \cite{Medina07,Tang07,Guan08,Mo09,Rong09,Rong2009TWC}. There are various design criteria with different goals. The most common criteria are capacity maximization \cite{Tang07,Medina07,Rong09} and data mean-square-error (MSE) minimization \cite{Guan08,Mo09,Rong09}. In most of the previous works on transceiver design, most of the designs are restricted for dual-hop relaying systems and furthermore channel state information (CSI) is assumed to be perfectly known. Unfortunately, channel estimation errors are inevitable in practical systems. To mitigate the effect on the performance of AF relaying systems, such channel estimation errors should be taken into account in the transceiver design process.




In this paper, we consider robust transceiver design for a \textbf{\textsl{multi-hop}} AF relaying system with channel estimation errors. Taking the Gaussian distributed channel errors into account, the precoder at source, multiple forwarding matrices at all the relays and equalizer at destination are jointly designed. The structure of the optimal solution for the unified problem is derived based on Majorization theory and properties of vector-monotone functions. The derived optimal structure covers most of the existing transceiver design results in point-to-point and dual-hop AF MIMO relaying systems as special cases. With the optimal structure, iterative water-filling solutions are proposed to obtain the remaining unknown parameters in the transceiver. The performance advantages of the proposed robust designs are demonstrated by simulation results.

The following notations are used throughout this paper. Boldface
lowercase letters denote vectors, while boldface uppercase letters
denote matrices. The notation ${\bf{Z}}^{\rm{H}}$ denotes the
Hermitian of the matrix ${\bf{Z}}$, and ${\rm{Tr}}({\bf{Z}})$ is the
trace of the matrix ${\bf{Z}}$. The symbol ${\bf{I}}_{M}$ denotes the
$M \times M$ identity matrix, while ${\bf{0}}_{M,N}$ denotes the $M
\times N$ all zero matrix. The notation ${\bf{Z}}^{1/2}$ is the
Hermitian square root of the positive semidefinite matrix
${\bf{Z}}$, such that ${\bf{Z}}^{1/2}{\bf{Z}}^{1/2}={\bf{Z}}$ and
${\bf{Z}}^{1/2}$ is also a Hermitian matrix. The symbol $\lambda_i({\bf{Z}})$ represents the $i^{\rm{th}}$ largest eigenvalue of ${\bf{Z}}$. The symbol $\otimes$ denotes
the Kronecker product. For two Hermitian matrices, ${\bf{C}} \succeq
{\bf{D}}$ means that ${\bf{C}}-{\bf{D}}$ is a positive semi-definite
matrix. For two vectors, ${\bf{x}}\ge {\bf{y}}$ represents that each element of ${\bf{x}}$ is larger than  the corresponding counterpart of ${\bf{y}}$. The symbol ${\boldsymbol \Lambda} \searrow $ represents a rectangular diagonal
matrix with decreasing diagonal elements.

\section{System Model}
\label{set: system model}

In this paper, a multi-hop AF MIMO relaying system is considered. There is one source with $N_1$ antennas wants to communicate with the destination with $M_{K}$ antennas through $K-1$ relays. For the $k^{\rm{th}}$ relay, it has $M_{k}$ receive
antennas and $N_{k+1}$ transmit antennas. It is obvious that the dual-hop AF MIMO relaying systems is one of its special cases when $K=2$.

At the source, a $N\times 1$ data vector s with  covariance matrix ${\bf{R}}_{\bf{s}} = {\mathbb{E}}\{{\bf{s}}{\bf{s}}^{\rm H}\} = {\bf{I}}_N$ is transmitted  through a precoder matrix ${\bf{P}}_1$. The received signal ${\bf{x}}_1$ at the first relay is
\begin{align}
{\bf{x}}_1= {\bf{H}}_{1}{\bf{P}}_1{\bf{s}}+{\bf{n}}_1
\end{align}where ${\bf{H}}_{1}$ is the MIMO channel matrix between the source and the first relay, and ${\bf{n}}_1$ is the additive Gaussian noise vector at the first relay with zero mean and
covariance matrix ${\bf{R}}_{n_1}=\sigma_1^2{\bf{I}}_{M_1}$.

At the first relay, the received signal ${\bf{x}}_1$ is first multiplied by a
forwarding matrix ${\bf{P}}_2$ and then the resultant signal is
transmitted to the second relay. The received signal ${\bf{x}}_2$ at
the second relay is given by
\begin{align}
\label{equ:signal} {\bf{x}}_2 &={\bf{H}}_{2}{\bf{P}}_2{\bf{x}}_1+{\bf{n}}_2 \nonumber \\
&={{\bf{H}}_{2} {\bf{P}}_2
{\bf{H}}_{1}{\bf{P}}_1{\bf{s}}}  + {{\bf{H}}_{2} {\bf{P}}_2{\bf{n}}_1
} + {\bf{n}}_2,
\end{align}where ${\bf{H}}_{2}$ is the MIMO channel matrix between the first relay
and the second relay, and ${\bf{n}}_2$ is the additive Gaussian noise
vector at the second relay with zero mean and covariance matrix ${\bf{R}}_{n_2}=\sigma_2^2{\bf{I}}_{M_2}$. Similarly, the received signal at $k^{\rm{th}}$ relay can be written as
\begin{align}
{\bf{x}}_{k}={\bf{H}}_{k}{\bf{P}}_{k}{\bf{x}}_{k-1}+{\bf{n}}_{k}
\end{align}where ${\bf{H}}_k$ is the channel for the $k^{\rm{th}}$ hop, and ${\bf{n}}_{k}$ is the additive Gaussian noise with zero mean and covariance matrix ${\bf{R}}_{n_{k}}=\sigma_{k}^2{\bf{I}}_{M_{k}}$.

Finally, for a $K$-hop AF MIMO relaying system, the received signal at the destination is
\begin{align}
{\bf{y}} = \left[{\prod_{k=1}^K}{\bf{H}}_{k}{\bf{P}}_k\right]{\bf{s}}  + \sum_{k=1}^{K-1}\left\{ \left[\prod_{l={k+1}}^K{\bf{H}}_{l}{\bf{P}}_l\right]{\bf{n}}_k
\right\}+{\bf{n}}_K,
\end{align}where ${\prod_{k=1}^K}{\bf{Z}}_k$ denotes ${\bf{Z}}_K\times \cdots \times {\bf{Z}}_1$. In order to guarantee the transmitted data ${\bf{s}}$ can be
recovered at the destination, it is assumed that
$N_k$ and $M_k$ are greater than or equal to $N$ \cite{Guan08}.

In practical systems, because of limited length of training sequences, channel estimation errors are inevitable.
With channel estimation errors, we can write
\begin{align}
 {\bf{H}}_{k}&={\bf{\bar H}}_{k}+\Delta{\bf{H}}_{k},
\end{align}where ${\bf{\bar H}}_{k}$ is the estimated channel in the $k^{\rm{th}}$ hop and $\Delta{\bf{H}}_{k}$ is the corresponding channel estimation error whose elements are zero mean Gaussian random variables. Moreover, the $M_k \times
N_k$ matrix $\Delta{\bf{H}}_{k}$ can be decomposed using the widely used Kronecker model
$\Delta{\bf{H}}_{k}=
{\boldsymbol{\Sigma}}_{k}^{{1}/{2}}{\bf{H}}_{W,k}
{\boldsymbol{\Psi}}_{k}^{{1}/{2}}$ \cite{Xing10,Xing1012}. The elements of the $M_k \times N_k$ matrix
${\bf{H}}_{W,k}$ are independent and identically distributed
(i.i.d.) Gaussian random variables with zero mean and unit variance. The specific formulas of the row correlation matrix ${\boldsymbol \Sigma}_k$ and the  column correlation matrix ${\boldsymbol \Psi}_k$ are determined by the training sequences and channel estimators being used \cite{Xing10,Xing1012}.

At the destination, a linear equalizer ${\bf{G}}$ is employed to
detect the desired data vector ${\bf{s}}$. The resulting data MSE
matrix equals to $
{\boldsymbol{\Phi}}({\bf{G}})=\mathbb{E}\{({\bf{G}}{\bf{y}}-{\bf{s}})({\bf{G}}{\bf{y}}-{\bf{s}})^{\rm{H}}\} $, where the expectation is taken with respect to random data, channel estimation errors, and noise.
Following a similar derivation in dual-hop systems \cite{Xing10}, the MSE matrix is derived to be
\begin{align}
\label{MSE_final} &{\boldsymbol {\Phi}}({\bf{G}})\nonumber \\
&=\mathbb{E}\{({\bf{G}}{\bf{y}}-{\bf{s}})
({\bf{G}}{\bf{y}}-{\bf{s}})^{\rm{H}}\}\nonumber \\& ={\bf{G}}[{\bf{\bar
H}}_{K}{\bf{P}}_K{\bf{R}}_{{\bf{x}}_{K-1}}{\bf{P}}_K^{\rm{H}} {\bf{\bar
H}}_{K}^{\rm{H}}+ {\rm{Tr}}({\bf{P}}_K{\bf{R}}_{{\bf{x}}_{K-1}} {\bf{P}}_K^{\rm{H}}
{\boldsymbol\Psi} _{K}
   ){\boldsymbol \Sigma}_{K} \nonumber \\
&+{\bf{R}}_{n_K}]{\bf{G}}^{\rm{H}}+ {\bf{I}}_N-\left[\prod_{k=1}^K{\bf{\bar H}}_{k}{\bf{P}}_k\right]^{\rm{H}}{\bf{G}}^{\rm{H}}-
{\bf{G}}\left[\prod_{k=1}^K{\bf{\bar H}}_{k}{\bf{P}}_k\right],
\end{align}where the received signal covariance matrix ${\bf{R}}_{{\bf{x}}_k}$ at the $k^{\rm{th}}$ relay satisfies the following recursive formula
\begin{align}
\label{R_x}
{\bf{R}}_{{\bf{x}}_k}&= {\bf{\bar H}}_{k}{\bf{P}}_k{\bf{R}}_{{\bf{x}}_{k-1}}{\bf{P}}_k^{\rm{H}}{\bf{\bar H}}_{k}^{\rm{H}}+ { {\rm{Tr}}({\bf{P}}_k{\bf{R}}_{{\bf{x}}_{k-1}} {\bf{P}}_k^{\rm{H}}
{\boldsymbol\Psi} _{k}
   ){\boldsymbol\Sigma} _{k}+{\bf{R}}_{n_k}},
\end{align}and ${\bf{R}}_{\bf{x}_0}={\bf{R}}_{\bf{s}}={\bf{I}}_N$ represents the signal covariance matrix at the source.

\section{Transceiver Design Problems}
\label{sect:problem formulation}

\subsection{Objective Functions}

There are various performance metrics for transceiver designs. In the following, we focus on two widely used metrics.

(1) In general, for balancing the performance across different data streams, (e.g., minimizing the worst data stream MSE), the objective function is written as \cite{Palomar03}
\begin{align}
\label{Obj 3}
\text{Obj 1:}\ &{\boldsymbol \psi}_{1}[{\rm{d}}({\boldsymbol \Phi}({\bf{G}}))]
\end{align}where ${\boldsymbol \psi}_{1}(\bullet)$ is an increasing Schur-convex function\footnote{The detailed introduction of Schur-concave/convext functions, and majorization theory is given in \cite{Marshall79}.} and ${\rm{d}}({\boldsymbol \Phi}({\bf{G}}))=\left[[{\boldsymbol \Phi}({\bf{G}})]_{1,1}\ [{\boldsymbol \Phi}({\bf{G}})]_{2,2} \ \cdots \right]^{\rm{T}}$, with the symbol $[{\bf{Z}}]_{i,j}$ represents the $(i,j)^{\rm{th}}$ entry of ${\bf{Z}}$.

(2) On the other hand, if a preference is given over a certain data streams, (e.g., loading more resources to the data streams with better channel state information), the objective function can be written as
\begin{align}
\label{Obj 4}
\text{Obj 2:}\ &{\boldsymbol \psi}_{2}[{\rm{d}}({\boldsymbol \Phi}({\bf{G}}))]
\end{align}where ${\boldsymbol \psi}_{2}(\bullet)$ is an increasing Schur-concave function.

\subsection{Problem Formulation}
Although the above two criteria aim at different designs, the transceiver design optimization problem can be unified into a single form:
\begin{align}
\label{Opt_X_000}
\ & \min_{{\bf{P}}_k,{\bf{G}}} \ \ \  {\boldsymbol f}({\boldsymbol \Phi}({\bf{G}})) \nonumber \\
& \  {\rm{s.t.}} \ \ \ \ {\rm{Tr}}({\bf{P}}_k{\bf{R}}_{{\bf{x}}_{k-1}}{\bf{P}}_k^{\rm{H}}) \le P_k, \ \ k=1,\cdots, K
\end{align}where the objective function ${\boldsymbol f}(\bullet)$ is a real-valued matrix-variate function with ${\boldsymbol \Phi}({\bf{G}})$ as its argument.  Notice that for all the two objectives described above, ${\boldsymbol f}(\bullet)$ is a matrix-monotone increasing function.

For (\ref{Opt_X_000}), there is no constraint on the equalizer ${\bf{G}}$. We can differentiate the trace of (\ref{MSE_final}) with respect to ${\bf{G}}$ and obtain the LMMSE equalizer
\begin{align}
\label{G} {\bf{G}}_{\rm{LMMSE}}& =\left[\prod_{k=1}^K{\bf{\bar H}}_{k}{\bf{P}}_k\right]^{\rm{H}}[{\bf{\bar
H}}_{K}{\bf{P}}_K{\bf{R}}_{{\bf{x}}_{K-1}}{\bf{P}}_K^{\rm{H}} {\bf{\bar
H}}_{K}^{\rm{H}}\nonumber \\
&+{ {\rm{Tr}}({\bf{P}}_K{\bf{R}}_{{\bf{x}}_{K-1}} {\bf{P}}_K^{\rm{H}}
{\boldsymbol\Psi} _{K}
   ){\boldsymbol\Sigma} _{K}+{\bf{R}}_{n_K}}]^{-1},
\end{align}with the property \cite{Kay93}
\begin{align}
\label{G_OPT}
{\boldsymbol \Phi}({\bf{G}}_{\rm{LMMSE}}) \preceq {\boldsymbol \Phi}({\bf{G}}).
\end{align} Because ${\boldsymbol f}(\bullet)$ is a matrix-monotone increasing function, (\ref{G_OPT}) implies that ${\bf{G}}_{\rm{LMMSE}}$ minimizes the objective function in (\ref{Opt_X_000}). Substituting the optimal equalizer of (\ref{G}) into ${\boldsymbol \Phi}({\bf{G}})$ in (\ref{MSE_final}), ${\boldsymbol \Phi}({\bf{G}})$ equals to
\begin{align}
\label{MSE_Matrix_Orig}
{\boldsymbol {\Phi}}_{\rm{MSE}}& ={\bf{I}}_N-\left[\prod_{k=1}^K{\bf{\bar H}}_{k}{\bf{P}}_k\right]^{\rm{H}}[{\bf{\bar
H}}_{K}{\bf{P}}_K{\bf{R}}_{{\bf{x}}_{K-1}}{\bf{P}}_K^{\rm{H}} {\bf{\bar
H}}_{K}^{\rm{H}}\nonumber \\
&+{ {\rm{Tr}}({\bf{P}}_K{\bf{R}}_{{\bf{x}}_{K-1}} {\bf{P}}_K^{\rm{H}}
{\boldsymbol\Psi} _{K}
   ){\boldsymbol\Sigma} _{K}+{\bf{R}}_{n_K}}]^{-1} \left[\prod_{k=1}^K{\bf{\bar H}}_{k}{\bf{P}}_k\right].
\end{align}

For multi-hop AF MIMO relaying systems, the received signal at $k^{\rm{th}}$ relay depends on the forwarding matrices at all preceding relays, making the power allocations at different relays couples with each other (as seen in the constrains of (\ref{Opt_X_000})), and thus the problem (\ref{Opt_X_000}) difficult to solve.
In order to simplify the problem, we define the following new variable in terms of ${\bf{P}}_k$:
\begin{align}
\label{F_definition}
{\bf{F}}_k&\triangleq
{\bf{P}}_k{\bf{K}}_{{\bf{F}}_{k-1}}^{1/2}\nonumber \\
&\times(\underbrace{{\bf{K}}_{{\bf{F}}_{k-1}}^{-1/2}{\bf{\bar
H}}_{k-1}{\bf{F}}_{k-1}{\bf{F}}_{k-1}^{\rm{H}}{\bf{\bar
H}}_{k-1}^{\rm{H}}{\bf{K}}_{{\bf{F}}_{k-1}}^{-1/2}+{\bf{I}}_{M_{k-1}}}_{\triangleq {\boldsymbol \Pi}_{k-1}})^{1/2}{\bf{Q}}_{k-1}^{\rm{H}},
\end{align}where ${\bf{K}}_{{\bf{F}}_k}\triangleq {\rm{Tr}}({\bf{F}}_k{\bf{F}}_k^{\rm{H}}
{\boldsymbol\Psi} _{k}
   ){\boldsymbol\Sigma} _{k}+{\sigma}_{n_k}^2{\bf{I}}_{M_k}$ and ${\bf{Q}}_k$ is an unknown unitary matrix. The introduction of ${\bf{Q}}_k$ is due to that fact
   that for a positive semi-definite matrix ${\bf{M}}$, its square roots has the form ${\bf{M}}^{1/2}{\bf{Q}}$ where ${\bf{Q}}$ is an unitary matrix. Notice that ${\bf{F}}_1={\bf{P}}_1$. With the new variable, the MSE matrix ${\boldsymbol { \Phi}}_{\rm{MSE}}$ is reformulated as
\begin{align}
\label{MSE_Matrix_M}
{\boldsymbol { \Phi}}_{\rm{MSE}}
 & ={\bf{I}}_N-\left[\prod_{k=1}^K{\bf{Q}}_k{\boldsymbol \Pi}_k^{-1/2}{\bf{K}}_{{\bf{F}}_k}^{-1/2}{\bf{\bar H}}_k{\bf{F}}_k\right]^{\rm{H}}\nonumber \\
 & \times \left[\prod_{k=1}^K\underbrace{{\bf{Q}}_k{\boldsymbol \Pi}_k^{-1/2}{\bf{K}}_{{\bf{F}}_k}^{-1/2}{\bf{\bar H}}_k{\bf{F}}_k}_{\triangleq  {\boldsymbol A}_{k}}\right]\nonumber \\
 &={\bf{I}}_N-{\boldsymbol A}_{1}^{\rm{H}}\cdots{\boldsymbol A}_{K}^{\rm{H}} {\boldsymbol A}_{K}\cdots{\boldsymbol A}_{1}.
\end{align} Meanwhile, with the new variables ${\bf{F}}_k$,  the corresponding power constraint in the $k^{\rm{th}}$ hop can now be rewritten as
\begin{align}
\label{constraint}
{\rm{Tr}}({\bf{F}}_k{\bf{F}}_k^{\rm{H}}) \le P_k.
\end{align}
 It is obvious that with the new variables ${\bf{F}}_k$, the constraints become independent of each other. Putting (\ref{MSE_Matrix_M}) and (\ref{constraint}) into (\ref{Opt_X_000}), the transceiver design problem can be reformulated as
\begin{align}
\label{Opt_X}
{\textbf{P 1:}} \ & \min_{{\bf{F}}_k,{\bf{Q}}_k} \ \ \  {\boldsymbol f}( {\bf{I}}_N-{\boldsymbol \Theta}) \nonumber \\
& \  {\rm{s.t.}} \ \ \ \ \ {\rm{Tr}}({\bf{F}}_k{\bf{F}}_k^{\rm{H}}) \le P_k, \ \ k=1,\cdots, K  \nonumber \\
& \ \ \ \ \ \   \  \ \ \  {\boldsymbol \Theta}={\boldsymbol A}_{1}^{\rm{H}}\cdots{\boldsymbol A}_{K}^{\rm{H}} {\boldsymbol A}_{K}\cdots{\boldsymbol A}_{1} \nonumber \\
& \ \ \ \ \ \   \  \ \ \ {\bf{Q}}_k^{\rm{H}}{\bf{Q}}_k={\bf{I}}_{M_k}
\end{align}

From the definition of ${\boldsymbol A}_k$ in (\ref{MSE_Matrix_M}) and noticing that ${\bf{K}}_{{\bf{F}}_k}={\rm{Tr}}({\bf{F}}_k{\bf{F}}_k^{\rm{H}}{\boldsymbol \Psi}_k){\boldsymbol \Sigma}_k+\sigma_{n_k}^2{\bf{I}}_{M_k}$, it can be seen that ${{\bf{F}}_k}$ appears at multiple positions in the objective function. Therefore,
the optimization problem is much more complicated than the counterpart with prefect CSI. Indeed, as demonstrated by existing works, robust transceiver design for point-to-point or dual-hop relaying MIMO systems is much more complicated and challenging than its counterpart with perfect CSI \cite{Xing1012,Xing10}.

%

\section{Optimal Structure of Robust Transceiver}
\label{sect:structure}

Based on the formulations of the objectives given in (\ref{Obj 3}) and (\ref{Obj 4}), in Appendix~\ref{Appedix:5}, it is proved that \textbf{P 1} has the following property.

\noindent \textbf{Property 1:} At the optimal value of \textbf{P 1}, ${\boldsymbol{\Theta}}$ must have the structure of
\begin{align}
{\boldsymbol \Theta}={\bf{U}}_{\boldsymbol \Omega}{\rm{diag}}[{\boldsymbol \lambda}({\boldsymbol \Theta})]{\bf{U}}_{\boldsymbol \Omega}^{\rm{H}}
\end{align}where the vector $
{\boldsymbol \lambda}({\boldsymbol \Theta})=[{\lambda}_1({\boldsymbol \Theta}), \cdots, {\lambda}_N({\boldsymbol \Theta})]^{\rm{T}}$ with ${\lambda}_n({\boldsymbol \Theta})$ being the $n^{\rm{th}}$ largest eigenvalue of ${\boldsymbol \Theta}$, and
\begin{align}
\label{U}
{\bf{U}}_{\boldsymbol{\Omega}}=\left\{ {\begin{array}{*{20}c}
   {{\bf{Q}}_{\bf{F}}\ \ \ \ \text{for Obj 1}}  \\
   { \ \ {\bf{I}}_N \ \ \  \text{for Obj 2}}  \\
\end{array}} \right ..
\end{align}In (\ref{U}), the unitary matrix ${\bf{U}}_{\bf{W}}$ is defined from the eigen-decomposition ${\bf{W}}={\bf{U}}_{\bf{W}}{\boldsymbol{\Lambda}}_{\bf{W}}{\bf{U}}_{\bf{W}}^{\rm{H}}$ with ${\boldsymbol{\Lambda}}_{\bf{W}}\searrow$, the matrix ${\bf{U}}_{\rm{Arb}}$ is an arbitrary unitary matrix, and ${\bf{Q}}_{\bf{F}}$ is the unitary matrix which makes ${\bf{Q}}_{\bf{F}}{\rm{diag}}[{\boldsymbol \lambda}({\boldsymbol \Theta})]{\bf{Q}}_{\bf{F}}^{\rm{H}}$ having identical diagonal elements. Furthermore, with this optimal structure, the objective function of \textbf{P 1} equals to
\begin{align}
&{\boldsymbol f}({\bf{I}}_N-\underbrace{{\bf{U}}_{\boldsymbol \Omega}{\rm{diag}}[{\boldsymbol \lambda}({\boldsymbol \Theta})]{\bf{U}}_{\boldsymbol \Omega}^{\rm{H}}}_{\boldsymbol \Theta})={\boldsymbol g}[{\boldsymbol \lambda}({\boldsymbol \Theta})]
\end{align}where ${\boldsymbol  g}(\bullet)$ is a monotobically decreasing and Schur-concave function with respective to ${\boldsymbol \lambda}({\boldsymbol \Theta})$. \footnote{The specific expressions of ${\boldsymbol g}(\bullet)$ are given in Appendix~\ref{Appedix:5}, but they are not important for the derivation of the optimal structures. }

\noindent \textbf{Proof:} See Appendix~\ref{Appedix:5}. $\blacksquare$

Based on \textbf{Property 1}, the objective function of (\ref{Opt_X}) can be directly replaced by ${\boldsymbol g} [{\boldsymbol \lambda}({\boldsymbol \Theta})]$ and thus the optimization problem is simplified as
\begin{align}
\label{P_1_2}
{\textbf{P 2:}} \  & \min_{{\bf{F}}_k,{\bf{Q}}_k} \ \ \  {\boldsymbol g} [{\boldsymbol \lambda}({\boldsymbol \Theta})] \nonumber \\
& \ \ {\rm{s.t.}} \ \ \ \  {\boldsymbol \Theta}={\boldsymbol A}_{1}^{\rm{H}}\cdots{\boldsymbol A}_{K}^{\rm{H}} {\boldsymbol A}_{K}\cdots{\boldsymbol A}_{1} \nonumber \\
& \ \  \ \ \ \ \ \ \ \ {\rm{Tr}}({\bf{F}}_k{\bf{F}}_k^{\rm{H}}) \le P_k, \ \ {\bf{Q}}_k^{\rm{H}}{\bf{Q}}_{k}={\bf{I}}_{M_k} \nonumber \\
&  \ \  \ \ \ \ \ \ \ \             {\boldsymbol \Theta}={\bf{U}}_{\boldsymbol \Omega}{\rm{diag}}[{\boldsymbol \lambda}({\boldsymbol \Theta})]{\bf{U}}_{\boldsymbol \Omega}^{\rm{H}}
\end{align}where ${\boldsymbol{A}}_k$'s are defined in (\ref{MSE_Matrix_M}). In order to further simplify the optimization problem, we make use of the following two additional properties.

\noindent \underline{\textsl{\textbf{Property 2:}}} As ${\boldsymbol g}(\bullet)$ is a decreasing and Schur-concave function and ${\boldsymbol \lambda}(\boldsymbol \Theta)\prec_w{\boldsymbol \gamma}(\boldsymbol \Theta)$, the objective function in \textbf{P 2} satisfies
\begin{align}
&{\boldsymbol g}({\boldsymbol \lambda}({\boldsymbol \Theta}))\ge{\boldsymbol g}( \underbrace{[{\gamma}_{1}(\boldsymbol \Theta) \ \cdots {\gamma}_{N}(\boldsymbol \Theta)]^{\rm{T}}}_{\triangleq {\boldsymbol {\gamma}}(\boldsymbol \Theta)} ) \label{inequ_3} \\
\text{with} \ \ & {\gamma}_{i}(\boldsymbol \Theta)\triangleq \lambda_i({\boldsymbol A}_{K}^{\rm{H}}{\boldsymbol A}_K)\lambda_i({\boldsymbol A}_{K-1}^{\rm{H}}{\boldsymbol A}_{K-1})\cdots \lambda_i({\boldsymbol A}_1^{\rm{H}}{\boldsymbol A}_1),
\end{align}where the equality in (\ref{inequ_3}) holds when the neighboring ${\boldsymbol{A}}_k$'s satisfy
\begin{align}
{\bf{V}}_{{\boldsymbol{A}}_k}={\bf{U}}_{{\boldsymbol{A}}_{k-1}}, \ \ k=2,\cdots,K
\end{align}with unitary matrices ${\bf{U}}_{{\boldsymbol{A}}_k}$ and ${\bf{V}}_{{\boldsymbol{A}}_{k}}$ being defined based on the following singular value decomposition ${\boldsymbol{A}}_k={\bf{U}}_{{\boldsymbol{A}}_k}
{\boldsymbol{\Lambda}}_{{\boldsymbol{A}}_k}{\bf{V}}_{{\boldsymbol{A}}_k}^{\rm{H}} \ \ \text{with} \  \ {\boldsymbol{\Lambda}}_{{\boldsymbol{A}}_k} \searrow$.


\noindent \textbf{\underline{\textsl{Property 3:}}} As ${\boldsymbol g}(\bullet)$ is a monotonically decreasing function with respective to its vector argument, the optimal solutions of the optimization problem  always occur on the boundary:
\begin{align}
\label{Power_constraint}
{\rm{Tr}}({\bf{F}}_{k}{\bf{F}}_{k}^{\rm{H}})=P_k.
\end{align}Furthermore, defining
\begin{align}
\label{eta_f}
& \eta_{f_k} \triangleq {\rm{Tr}}({\bf{F}}_k{\bf{F}}_k^{\rm{H}}{\boldsymbol \Psi}_k)\alpha_k+\sigma_{n_k}^2
\end{align}with $\alpha_k={\rm{Tr}}({\boldsymbol \Sigma}_k)/M_k$ which is a constant, (\ref{Power_constraint}) is  equivalent to
\begin{align}
{\rm{Tr}}[{\bf{F}}_k
{\bf{F}}_k^{\rm{H}}(\alpha_k P_k{\boldsymbol \Psi}_k+\sigma_{n_k}^2{\bf{I}}_{N_k})]/\eta_{f_k}=P_k.
\end{align}

Based on \textbf{Properties 2} and \textbf{3} , the optimal solution of the optimization problem (\ref{P_1_2}) is exactly the optimal solution of the following new optimization problem with different constraints
\begin{align}
\label{Opt_X_Final}
{\textbf{P 3:}} \ & \min_{{\bf{F}}_k,{\bf{Q}}_k} \ \ \  {\boldsymbol g} [{\boldsymbol \gamma}({\boldsymbol \Theta})]\nonumber \\
& \ \ \ {\rm{s.t.}} \ \ \ \  {\rm{Tr}}[{\bf{F}}_{k}
{\bf{F}}_{k}^{\rm{H}}(\alpha_kP_k{\boldsymbol \Psi}_k+\sigma_{n_k}^2{\bf{I}}_{N_k})]/\eta_{f_k}=P_k \nonumber \\ &  \ \ \ \ \ \ \ \ \ \ \ {\boldsymbol \Theta}={\boldsymbol A}_{1}^{\rm{H}}\cdots{\boldsymbol A}_{K}^{\rm{H}} {\boldsymbol A}_{K}\cdots{\boldsymbol A}_{1} \nonumber \\
 & \ \ \ \ \ \ \ \ \ \ \  {\bf{Q}}_{k}^{\rm{H}}{\bf{Q}}_k={\bf{I}}_{M_k},  \ \           {\boldsymbol \Theta}={\bf{U}}_{\boldsymbol \Omega}{\rm{diag}}[{\boldsymbol \gamma}({\boldsymbol \Theta})]{\bf{U}}_{\boldsymbol \Omega}^{\rm{H}}\nonumber \\
 & \ \ \ \ \ \ \ \ \ \ \ {\bf{V}}_{{\boldsymbol{A}}_k}={\bf{U}}_{{\boldsymbol{A}}_{k-1}}, \ \ k=2,\cdots,K.
\end{align}Noticing that $ {\boldsymbol g} (\bullet)$ is a monotonically decreasing function, solving \textbf{P 3} gives the following structure for the optimal solution.

\noindent \textbf{\underline{\textsl{Conclusion 1:}}} Defining  unitary matrices ${\bf{U}}_{{\boldsymbol{\mathcal{H}}}_k}$ and ${\bf{V}}_{{\boldsymbol{\mathcal{H}}}_k}$ based on the following singular value decomposition
\begin{align}
\label{decomp}
& ({\bf{K}}_{{\bf{F}}_k}/\eta_{f_k})^{-1/2}{\bf{\bar H}}_k(\alpha_k P_k{\boldsymbol \Psi}_k+{\sigma}_{n_k}^2{\bf{I}}_{N_k})^{-1/2}={\bf{U}}_{{\boldsymbol{\mathcal{H}}}_k}{\boldsymbol \Lambda}_{{\boldsymbol{\mathcal{H}}}_k}{\bf{V}}_{{\boldsymbol{\mathcal{H}}}_k}^{\rm{H}}\nonumber \\
&  \text{with} \ \  {\boldsymbol \Lambda}_{{\boldsymbol{\mathcal{H}}}_k} \searrow \ \ \text{and} \ \ {\bf{U}}_{{\boldsymbol{\mathcal{H}}}_0}={\bf{U}}_{\boldsymbol{\Omega}},
\end{align}when ${\boldsymbol{ \Psi}}_k\propto {\bf{I}}$ or ${\boldsymbol \Sigma}_k\propto {\bf{I}}$, the optimal solutions of the optimization problem (\ref{Opt_X_Final}) have the following structure
\begin{align}
\label{Structure_F}
& {\bf{F}}_{k,\rm{opt}}=\sqrt{{\boldsymbol \xi}_k({\boldsymbol { \Lambda}}_{{\boldsymbol{\mathcal{F}}}_k})}(\alpha_k P_k{\boldsymbol
\Psi}_k+\sigma_{n_k}^2{\bf{I}}_{N_k})^{-1/2}\nonumber \\
& \ \ \ \ \ \ \ \ \ \ \ \ \ \ \times {\bf{V}}_{{\boldsymbol{\mathcal{H}}}_k,N}
{\boldsymbol { \Lambda}}_{{\boldsymbol{\mathcal{F}}}_k}{\bf{U}}_{{\boldsymbol{\mathcal{H}}}_{k-1},N}^{\rm{H}} \nonumber \\
&
{\bf{Q}}_{k,{\rm{opt}}}={\bf{I}}_{M_k},
\end{align}where ${\bf{V}}_{{\boldsymbol{\mathcal{H}}}_k,N}$ and ${\bf{U}}_{{\boldsymbol{\mathcal{H}}}_k,N}$ are the matrices consisting of the first $N$ columns of  ${\bf{V}}_{{\boldsymbol{\mathcal{H}}}_k}$ and ${\bf{U}}_{{\boldsymbol{\mathcal{H}}}_k}$, respectively, and ${\boldsymbol{ \Lambda}}_{{\boldsymbol{\mathcal{F}}}_k}$ is a $N \times N$ unknown diagonal matrix. The scalar ${\boldsymbol \xi}_k({\boldsymbol { \Lambda}}_{{\boldsymbol{\mathcal{F}}}_k})$ is a function of ${\boldsymbol{ \Lambda}}_{{\boldsymbol{\mathcal{F}}}_k}$and equals to
\begin{align}
{\boldsymbol \xi}_k({\boldsymbol { \Lambda}}_{{\boldsymbol{\mathcal{F}}}_k})&=\eta_{f_k}\nonumber \\&={\sigma_{n_k}^2}/\{1-\alpha_k {\rm{Tr}}[{\bf{V}}_{{\boldsymbol{\mathcal{H}}}_k,N}^{\rm{H}}(\alpha_k P_k{\boldsymbol
\Psi}_k+\sigma_{n_k}^2{\bf{I}}_{N_k})^{-1/2}\nonumber \\
& \times{\boldsymbol
\Psi}_k(\alpha_k P_k{\boldsymbol
\Psi}_k+\sigma_{n_k}^2{\bf{I}}_{N_k})^{-1/2}
{\bf{V}}_{{\boldsymbol{\mathcal{H}}}_k,N}{\boldsymbol{ \Lambda}}_{{\boldsymbol{\mathcal{F}}}_k}^2]\}.
\end{align}

In the optimal structure given by (\ref{Structure_F}), the scalar variable  ${\boldsymbol \xi}_k({\boldsymbol { \Lambda}}_{{\boldsymbol{\mathcal{F}}}_k})$ is only a function of the matrix ${\boldsymbol { \Lambda}}_{{\boldsymbol{\mathcal{F}}}_k}$ and therefore the only unknown variable in (\ref{Structure_F}) is ${\boldsymbol { \Lambda}}_{{\boldsymbol{\mathcal{F}}}_k}$. The remaining unknown diagonal elements of ${\boldsymbol { \Lambda}}_{{\boldsymbol{\mathcal{F}}}_k}$ can be obtained by water-filling alike solution as discussed in the next section.

\section{Computations of ${\boldsymbol { \Lambda}}_{{\boldsymbol{\mathcal{F}}}_k}$}
\label{sect:application}

The remaining unknown variables in (\ref{Structure_F}) are only ${\boldsymbol { \Lambda}}_{{\boldsymbol{\mathcal{F}}}_k}$. Substituting the optimal structures given by \textbf{Conclusion 1} into \textbf{P 3} and defining
$[{\boldsymbol {\Lambda}}_{{\boldsymbol{\mathcal{H}}}_k}]_{i,i}={h}_{k,i}$ and
$[{\boldsymbol { \Lambda}}_{{\boldsymbol{\mathcal{F}}}_k}]_{i,i}=f_{k,i}$ for $i=1,\cdots, N$, the optimization problem for computing ${\boldsymbol { \Lambda}}_{{\boldsymbol{\mathcal{F}}}_k}$ becomes
\begin{align}
\label{Opt_F}
 & \min_{{f}_{k,i}} \ \ \  {\boldsymbol g} [{\boldsymbol \gamma}({\boldsymbol \Theta})]\nonumber \\
& \ {\rm{s.t.}} \ \ \ \  \sum_{i=1}^N f_{k,i}^2=P_k \nonumber  \\
& \ \ \ \ \ \ \ \ \ {\boldsymbol \gamma}({\boldsymbol \Theta})=[{\gamma}_{1}({\boldsymbol \Theta}) \ \cdots {\gamma}_{N}({\boldsymbol \Theta})]^{\rm{T}}
\nonumber \\ &  \ \ \ \ \ \ \ \ \ { \gamma}_{i}({\boldsymbol \Theta})=\frac{{\prod_{k=1}^K}f_{k,i}^2h_{k,i}^2}{\prod_{k=1}^K(f_{k,i}^2h_{k,i}^2+1)}.
\end{align}The specific methods for finding $f_{k,i}$ depend on the expressions of ${\boldsymbol g} (\bullet)$, In the following, we discuss the solution of (\ref{Opt_F}) in more detail. The design criterion of MAX-MSE minimization is taken as example to show how to compute ${\boldsymbol { \Lambda}}_{{\boldsymbol{\mathcal{F}}}_k}$.

MAX-MSE minimization is a special case of \textbf{Obj 1} in (\ref{Obj 3}) and in this case,
${\boldsymbol \psi}_1({\rm{d}}({\boldsymbol \Phi}_{\rm{MSE}}))=\max [{\boldsymbol \Phi}_{\rm{MSE}}]_{i,i}$. Furthermore, in Appendix~\ref{Appedix:5} it is proved that ${\boldsymbol g} ({\boldsymbol \lambda}({\boldsymbol \Theta}))={\boldsymbol \psi}_1 [{\bf{1}}_N-({\sum}_{i=1}^N{\lambda}_{i}({\boldsymbol \Theta})/N )\otimes{\bf{1}}_N]$. Therefore, ${\boldsymbol g} [{\boldsymbol \gamma}({\boldsymbol \Theta})]$ equals to
\begin{align}
{\boldsymbol g}[{\boldsymbol \gamma}({\boldsymbol \Theta})]&=\max \left({\bf{1}}_N-({\sum}_{i=1}^N{\gamma}_{i}({\boldsymbol \Theta})/N )\otimes{\bf{1}}_N \right)\nonumber \\
&= 1-\frac{1}{N}\sum_{i=1}^N{\gamma}_{i}({\boldsymbol \Theta})
\end{align}based on which the optimization problem (\ref{Opt_F}) becomes
\begin{align}
\label{opt_scalar_MAX_MSE}
&\min_{f_{k,i}} \ 1-\frac{1}{N}\sum_{i=1}^N\frac{\prod_{k=1}^K (f_{k,i}^2h_{k,i}^2)}
{\prod_{k=1}^K (f_{k,i}^2h_{k,i}^2+1)} \nonumber \\
 &\ {\rm{s.t.}} \ \   \sum_{i=1}^N f_{k,i}^2 =P_k.
\end{align} The optimization problem (\ref{opt_scalar_MAX_MSE}) can be solved by using iterative water-filling algorithm.


\section{Simulation Results and Discussions}
\label{sect:simulation}
In this section, the performance of the proposed robust designs are evaluated by simulations. For the purpose of
comparison, the algorithms based on the estimated channel only
(without taking the channel estimation errors into account) are also simulated. In the following, we consider a three-hop AF MIMO relaying
system where all nodes are equipped with 4 antennas.
Furthermore, the estimation error
correlation matrices are chosen as the popular exponential model  \cite{Xing10} i.e., $[{\boldsymbol{\Psi}}_{k}]_{i,j}=\alpha^{|i-j|}$
and  $[{\boldsymbol{\Sigma}}_{k}]_{i,j}=\beta^{|i-j|}$.
The estimated
channels ${\bf{\bar H}}_{k}$'s, are
generated based on the following complex Gaussian distributions
\begin{align}
 &{\bf{\bar H}}_{k}\sim
\mathcal {C}\mathcal
{N}_{M_k,N_k}({\bf{0}}_{M_k,N_k},\frac{(1-\sigma_{e}^2)}{\sigma_{e}^2}{\boldsymbol
\Sigma}_{k}
\otimes {\boldsymbol \Psi}_{k}^{\rm{T}}),
\end{align} such that channel realizations ${\bf{H}}_{k}={\bf{\bar
H}}_{k}+\Delta{\bf{H}}_{k}$ have unit variance. We define the signal-to-noise ratio (${\rm{SNR}}$) for the $k^{\rm{th}}$ link
 as $P_k/\sigma_{n_k}^2$. At the source node, four independent data streams are transmitted and in each data stream, \begin{figure}[!ht] \centering
\includegraphics[width=.4\textwidth]{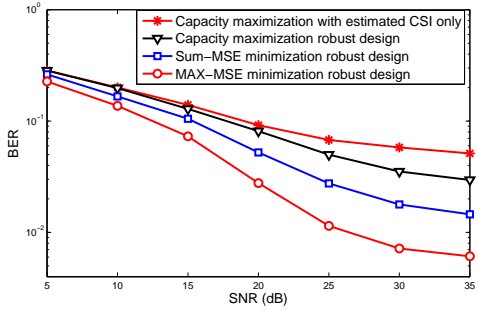}
\caption{BERs of the proposed robust design with different design objectives, when $\alpha=0.4$, $\beta=0$ and $\sigma_e^2=0.004$.}\label{fig:5}
\end{figure} ${N_{Data}}=10000$ independent QPSK symbols
are transmitted. Each point in the
following figure is an average of 10000 trials.
%
%
%
%
%
%

Fig.~\ref{fig:5} shows the bit error rate (BER) of the proposed robust designs  with different performance metrics: sum MSE minimization, mutual information maximization and MAX-MSE minimization. Other parameters are taken as $\alpha=0.4$, $\beta=0$ and $\sigma_{e}^2=0.004$. It can be seen that the former two criteria have better performance than the latter one.  Meanwhile, the capacity maximization based on estimated CSI only is given to show that the proposed robust designs are better than that of the design with estimated CSI only.

\section{Conclusions}
\label{sect:conclusion}
Bayesian robust transceiver design for multi-hop AF MIMO relaying systems under channel estimation errors was considered. Various transceiver design criteria were unified into a single optimization framework. Using majorization theory and properties of matrix-variate functions, the optimal structure of transceivers was derived. Then, the transceiver design problems were greatly simplified and the remaining unknowns were obtained by iterative water-filling solutions. The performance of the proposed transceiver designs has been demonstrated by simulation results.

\appendices

\section{Proof of Property 1}
\label{Appedix:5}

\noindent \underline{\textbf{Obj 1:}} For the diagonal elements of the positive semi-definite matrix ${\boldsymbol \Phi}_{\rm{MSE}}={\bf{I}}_N-{\boldsymbol \Theta}$, we have the following relationship \cite{Marshall79} \begin{align}
\label{APP_64}
{\bf{1}}_N-({\sum}_{i=1}^N{\lambda}_i({\boldsymbol \Theta})/N )\otimes{\bf{1}}_N \prec{\rm{d}}({\bf{I}}_N-{\boldsymbol \Theta})
\end{align}with the equality holds if and only if $[{\boldsymbol \Theta}]_{i,i}={\sum}_{i=1}^N{\lambda}_i({\boldsymbol \Theta})/N $, where ${\bf{1}}_N$ is the $N \times 1$ all one vector.

For the first objective function in (\ref{Obj 3}), as ${\boldsymbol \psi}_1(\bullet)$ is decreasing and Schur-convex, the objective function satisfies \cite{Palomar03}
\begin{align}
\label{App_Inequ_3}
& {\boldsymbol \psi}_1({\rm{d}}({\bf{I}}_N-{\boldsymbol \Theta}))\ge\underbrace{{\boldsymbol \psi}_1\left({\bf{1}}_N-({\sum}_{i=1}^N{\lambda}_i({\boldsymbol \Theta})/N )\otimes{\bf{1}}_N\right)}_{\triangleq{\boldsymbol g}[{\boldsymbol{\lambda}}({\boldsymbol \Theta})]},
\end{align}with equality holds if and only if $[{\boldsymbol \Theta}]_{i,i}={\sum}_{i=1}^N{\lambda}_i({\boldsymbol \Theta})/N $. Therefore, ${\boldsymbol \Theta}$ must have the following structure \cite{Palomar03}
\begin{align}
{\boldsymbol \Theta}={\bf{Q}}_{\bf{F}}{\rm{diag}}({\boldsymbol \lambda}({\boldsymbol \Theta})){\bf{Q}}_{\bf{F}}^{\rm{H}}.
\end{align}where ${\bf{Q}}_{\bf{F}}$ is a unitary matrix such that ${\boldsymbol \Theta}$ has identical diagonal elements.

Based on the definition that ${\boldsymbol \psi}_1(\bullet)$ is a decreasing and Schur-convex function, based on \textbf{A.6.Lemma} and \textbf{A.8.Lemma} in \cite{Marshall79} it can be directly proved that ${\boldsymbol g}({\boldsymbol{\lambda}}({\boldsymbol \Theta}))$ is a decreasing and Schur-concave function of ${\boldsymbol{\lambda}}({\boldsymbol \Theta})$.

\noindent \underline{\textbf{Obj 2:}} Notice that for the positive semi-definite matrix ${\boldsymbol \Phi}_{\rm{MSE}}={\bf{I}}_N-{\boldsymbol \Theta}$, ${\rm{d}}({\bf{I}}_N-{\boldsymbol \Theta}) \prec {\boldsymbol{\lambda}}({\bf{I}}_N-{\boldsymbol \Theta})$ \cite{Palomar03}. Furthermore ${\boldsymbol \psi}_2(\bullet)$ is Schur-concave, we have
\begin{align}
\label{App_Inequ_4}
{\boldsymbol \psi}_2({\rm{d}}({\bf{I}}_N-{\boldsymbol \Theta}))\ge \underbrace{{\boldsymbol \psi}_2({\bf{1}}_N-{\boldsymbol{\lambda}}({\boldsymbol \Theta}))}_{\triangleq {\boldsymbol g}[{\boldsymbol{\lambda}}({\boldsymbol \Theta})]}.
\end{align}In order to make the equality in (\ref{App_Inequ_4}) hold, we need $[{\boldsymbol \Theta}]_{i,i}={ \lambda}_i({\boldsymbol \Theta})$, which means that ${\boldsymbol \Theta}$ is a diagonal matrix. Therefore, we can write \begin{align}
{\boldsymbol \Theta}={\bf{I}}_N{\rm{diag}}({\boldsymbol \lambda}({\boldsymbol \Theta})){\bf{I}}_N.
\end{align}

Since ${\boldsymbol \psi}_2(\bullet)$ is increasing and Schur-concave,   based on \textbf{A.6.Lemma} and \textbf{A.8.Lemma} in \cite{Marshall79} it is obvious that ${\boldsymbol \psi}_2({\bf{1}}_N-{\boldsymbol{\lambda}}({\boldsymbol \Theta}))$ is decreasing and Schur-concave with respective to ${\boldsymbol{\lambda}}({\boldsymbol \Theta})$.

\end{document}